\documentclass{aa}

\usepackage{graphicx,subfigure,bm}
\usepackage{natbib}
\usepackage{txfonts}
\begin{document}

\title{Magnetic effect on equilibrium tides and its influence on the orbital evolution of binary systems}
\author{Xing Wei}
\institute{Department of Astronomy, Beijing Normal University, China \\ Email: xingwei@bnu.edu.cn}
\date{}
\titlerunning{Magnetic tide}
\authorrunning{Xing Wei}

\abstract{In the standard theory of equilibrium tides, hydrodynamic turbulence is considered. In this paper we study the effect of magnetic fields on equilibrium tides. We find that the turbulent Ohmic dissipation associated with a tidal flow is much stronger than the turbulent viscous dissipation such that a magnetic field can greatly speed up the tidal evolution of a binary system. We then apply the theory to  three binary systems:  the orbital migration of 51 Pegasi b, the orbital decay of WASP-12b, and the circularization of close binary stars. Theoretical predictions are in good agreement with  observations, which cannot be clearly interpreted with  hydrodynamic equilibrium tides.}
\maketitle

\section{Motivation}
Tides play a key role in the orbital evolution of a binary system, for example planet--satellite, star--planet, binary stars, and a compact object and its companion. The body on which a tide is exerted is called the primary and the body that raises a tide is called the secondary. Tidal force is the gravitational perturbation, which is inversely proportional to the cube of the orbital semimajor axis. A tide can be decomposed into two parts; one is the equilibrium tide, namely a large-scale deformation due to the hydrostatic balance between tidal potential and pressure perturbation, and the other is the dynamical tide, namely fluid waves (gravity waves and inertial waves) excited by tidal force. Equilibrium tides  dissipate through turbulence in the  convection zone, and the dynamical tides dissipate either through radiative cooling in the radiation zone (gravity wave) or through viscous shears in the convection zone (inertial wave).

The study of tides has a long history,  dating back to Newton and Laplace. In the 1950s  pioneering work on tidal theory began, and we  started to understand the orbital evolution of the satellites in the Solar System \citep{jeffreys1961, goldreich1963, goldreich1966} and that of the close binary stars in open clusters \citep{kopal1959, zahn1975, goldreich1977, zahn1977, hut1981, savonije1983, goldreich1989, zahn1989}. Since the 1990s with the detection of exoplanets, especially hot Jupiters (exoplanets close to their host stars with   orbital periods of several days), the theory and application of equilibrium tides were greatly developed \citep{lin1996, goodman1997, penev2009, barker2020, terquem2021}. However, the standard theory of equilibrium tides cannot provide sufficient tidal dissipation for the orbital evolution of either star--planet systems or binary star systems. Researchers then turned to dynamical tides \citep{lai1997, goodman1998, savonije2002, ogilvie2004, wu2005a, goodman2009, fuller2014} and  the nonlinear resonance locking of dynamical tides \citep{weinberg2012, wu2021}.

On the other hand, magnetic fields are ubiquitous in the Universe. In all the models mentioned above the magnetic effect on tides is not involved. In \citet{wei2016}, \citet{wei2018}, and \citet{ogilvie2018} the magnetic effect on dynamical tides was considered and it was found that magnetic fields can greatly enhance tidal dissipation. In this paper we consider the magnetic effect on equilibrium tides. We  find that magnetic fields can induce much stronger tidal dissipation, and thus a very significant effect on the orbital evolution of binary systems. In \S\ref{sec:formulation} we formulate the expressions of tidal dissipation and torques arising from hydrodynamical and magnetic turbulence. In \S\ref{sec:application} we apply our model to interpret the observations of the three binary systems. In \S\ref{sec:summary} a brief summary is given.

\section{Two tidal dissipations and torques}\label{sec:formulation}
We consider the major quadrupolar component of tidal potential resulting from orbital motion at frequency $\omega_o$ relative to the primary rotation at frequency $\omega_s$, namely an asynchronized tide. The tidal torque $\Gamma$ exerted on an orbit is related to tidal dissipation $D$ via $\Gamma = -D/(\omega_o-\omega_s)$. Either hydrodynamic or magnetic turbulence can lead to tidal torque, and the only difference lies in dissipation $D$, whether it is viscous dissipation $D_h$ or Ohmic dissipation $D_m$. Next we   find $D_h$ and $D_m$. We adopt the conventional notations: primary mass $M$, secondary mass $m$, primary radius $R$, and orbital semimajor axis $a$.

For hydrodynamic turbulence in the convection zone of a primary, viscous dissipation $D_h$ associated with the large-scale sheared tidal flow $u$ is
\begin{equation}\label{D_h}
D_h \approx \int\rho\nu_t \left(\frac{u}{d}\right)^2dV
\end{equation}
where $\rho$ is density, $\nu_t$ turbulent viscosity, $d$ thickness, and $V$ volume of the  convection zone. The magnitude of the tidal flow $u$ can be estimated with an equilibrium tide $\xi$ (i.e., $u=\dot\xi=2|\omega_o-\omega_s|\xi$), and the magnitude of $\xi$ can be estimated with the magnitude of tidal potential $\Psi=(Gmr^2/a^3)$ and self-gravity $g=GM(r)/r^2$ (i.e., $\xi=\Psi/g=(m/M(r))(r^4/a^3)$), where $r$ is the radius from the primary center and $M(r)$ is the mass enclosed within $r$. We consider a primary to be  a solar-type star with the base of convection zone at $r=R_c$. With the expressions of $\xi$, $u$, and $D_h$, we obtain the tidal torque due to hydrodynamic turbulence
\begin{equation}\label{Gamma_h}
\Gamma_h \approx -\beta_h\frac{m^2}{M}\left(\frac{R}{a}\right)^6 \nu_t(\omega_o-\omega_s).
\end{equation}
In the derivation, the mean density $\bar\rho=3M/4\pi R^3$ is employed, and the  structure factor $\beta_h=(4/3)/(1-\alpha)^2\int_\alpha^1 [x^{10}f/(\int_0^x x'^2fdx')^2]dx$ depending on density profile is introduced, where $\alpha=R_c/R$, $x=r/R$ and $f(x)=\rho(r)/\bar\rho$. By using the  MESA code \citep{Paxton2011} for the solar model, we readily obtain the structure factor $\beta_h\approx 0.224$ for the present Sun with $\alpha=0.7$. In Eq.  \eqref{Gamma_h}, if we replace $\nu_t$ with $R^2/t_f$, where $t_f$ is the friction timescale defined by Zahn in \citet{zahn1977}, then our torque expression \eqref{Gamma_h} is equivalent to Zahn's, namely Eq. (11) in \citet{zahn1989}, since $\beta_h$ and the apsidal constant $k_2$ are at the same order. On the other hand, in the next paragraph we show  that turbulent viscosity $\nu_t$ depends on position $r$, but in our order-of-magnitude estimation we neglect this dependence.

In the standard tidal theory, the magnitude of $\nu_t(\omega_o-\omega_s)$ in \eqref{Gamma_h} is  modeled with the tidal $Q$ parameter:  $Q=(GM/R)/(\nu_t|\omega_o-\omega_s|)$. Then  the key issue is how to estimate the turbulent viscosity $\nu_t$ or equivalently tidal $Q$. The mixing length theory provides an estimation $\nu_t\approx(1/3)v_c l$, where $v_c$ is the convective velocity of the largest turbulent eddies and $l$ is mixing length on which turbulent eddies complete momentum transport. The mixing length in the stellar convection zone is approximately twice the pressure scale height,  $l\approx 2p/(dp/dr)=2(\mathcal{R}/\mu_m)T/g$, where the hydrostatic balance $dp/dr=\rho g$ and the equation of state for ideal gas $p=(\mathcal{R}/\mu_m)\rho T$ are employed ($\mathcal{R}$ is gas constant and $\mu_m$ is mean molecular weight). The mixing length in the  stellar convection zone can then be estimated roughly at the order of 1/10 the thickness of convection zone. To estimate $v_c$, we assume that the work done by the buoyancy force on the mixing length is converted to kinetic energy,  $|\delta\rho|gl = (1/2)\rho v_c^2$, where $\delta\rho$ is the density deviation from the surroundings. On the other hand, the convective heat flux is proportional to convective velocity:  $F=\rho c_p(\delta T)v_c$. Combining these two expressions with the  thermodynamic relation $(\delta\rho)/\rho=-(\delta T)/T$ for ideal gas, we obtain the estimation $v_c = (2Fgl/\rho c_pT)^{1/3}\approx (F/\rho)^{1/3}$, where $l\approx 2(\mathcal{R}/\mu_m)T/g$ and $c_p=2.5\mathcal{R}/\mu_m$ are used. This estimation $v_c\approx(F/\rho)^{1/3}$ has been validated by numerical simulations \citep{chan1996, cai2014}. Taking the Sun as an  example, the average $v_c\approx(F/\bar\rho)^{1/3}\approx 35$ m/s which is comparable to the result of asteroseismology \citep{hanasoge2012}, $l\approx 10^7$ m, and therefore $\nu_t\approx(1/3)v_c l\approx 10^8$ m$^2$/s which is already validated by the numerical simulations \citep{cai2014, brandenburg2016}. In the standard tidal theory with weak friction assumption, turbulent viscosity $\nu_t$ does not depend on tidal frequency.

However, fast tides can suppress turbulent viscosity. There are two long-standing   arguments under debate about this suppression effect. One is the linear law based on the argument that turbulent transport occurs only on half of the tidal period \citep{zahn1977}, and the other is the square law based on the argument that only an eddy whose turnover timescale is faster than the tide can contribute to turbulent viscosity \citep{goldreich1977}. The earlier numerical simulations support the linear law \citep{penev2007, penev2009}, but the most recent numerical simulations support the square law \citep{barker2020a, barker2020b}. In additional to these two laws, the perburbative analysis yields the 5/3 law \citep{goodman1997}. We summarize the three laws here for the fast-tide suppression effect,
\begin{strip}
\begin{equation}\label{nu}
\nu_t\approx
\frac{1}{3}v_cl\cdot\min\left[\left(\frac{P_{tide}}{2\tau_c}\right),1\right], \hspace{3mm}
\frac{1}{3}v_cl\cdot\min\left[\left(\frac{P_{tide}}{2\pi\tau_c}\right)^2,1\right], \hspace{3mm}
v_cl\cdot\min\left[\left(\frac{P_{tide}}{2\pi\tau_c}\right)^{5/3},1\right]
\end{equation}
\end{strip}
where $P_{tide}=2\pi/(2|\omega_o-\omega_s|)$ is the tidal period and $\tau_c=l/v_c$ is the turnover timescale of the largest turbulent eddies.

After revisiting the equilibrium tide due to hydrodynamic turbulence, we now turn to magnetic turbulence. Ohmic dissipation associated with the large-scale tidal flow $u$ is
\begin{equation}\label{D_m}
D_m \approx \int\frac{J^2}{\sigma_t}dV \approx \int\frac{B^2}{\mu}\frac{u^2}{\eta_t}dV
,\end{equation}
where $\sigma_t$ is the turbulent electric conductivity, $\mu$ the magnetic permeability, $\eta_t=1/(\sigma_t\mu)$ the turbulent magnetic diffusivity, $B$ the mean magnetic field, and $J\approx\sigma_t u B$ the electric current induced by tidal flow. In Appendix A we show more details about why Ohmic dissipation is $\sim J^2/\sigma_t$ through the magnetic induction equation, namely the first ``$\approx$'' in Eq. \eqref{D_m}, and why the electric current $J$ associated with tide is $\sim\sigma_t u B$ through the order analysis in Ohm's law, namely the second ``$\approx$'' in \eqref{D_m}. To understand Eq. \eqref{D_m}, we consider the primary as a conductor. In the primary, the Ohmic dissipation rate = voltage$^2$/resistance. Voltage, which is the tidal flow times the magnetic field times the typical length scale, is fixed with the fixed external forcing (i.e., tidal flow). Resistance is proportional to turbulent magnetic diffusivity. Therefore, with the external tide fixed, lower turbulent magnetic diffusivity leads to lower resistance and hence higher Ohmic dissipation.

The surface field of the present Sun is roughly 10 gauss. At a young age, the surface field reaches 1000 gauss \citep{yu2017}. A star at a young age rotates at a period of several days and then spins down due to the loss of angular momentum via stellar wind and magnetic braking. Observations show that rapid rotation corresponds to a strong magnetic field \citep{wright2011, reiners2014}, and this is because rapidly rotating turbulence is anisotropic with columnar structure to induce dynamo action quite differently to a slow rotator \citep{davidson2013, wei2022}. The internal field is stronger than the surface field, and we take the ratio 3.5 \citep{christensen2009}. 

Turbulent magnetic diffusivity $\eta_t$ is comparable to turbulent viscosity $\nu_t$ since the two diffusivities are caused by the same physical process, namely turbulent transport, and this has  already been validated by observational constraints \citep{choudhuri2007} and numerical simulations \citep{brandenburg2003, brandenburg2020}. As we  already know, a fast tide suppresses turbulent viscosity $\nu_t$, and hence turbulent magnetic diffusivity $\eta_t$ as well. According to Eq. \eqref{D_m},  fast tides enhance tidal dissipation in magnetic turbulence, which is opposite to the standard theory of equilibrium tides in hydrodynamic turbulence. 

Comparing Ohmic dissipation \eqref{D_m} to viscous dissipation \eqref{D_h}, we find that the ratio of two tidal torques is roughly $\Gamma_m/\Gamma_h=D_m/D_h\approx(B^2/\mu)d^2/(\bar\rho\nu_t\eta_t)$. The ratio does not depend on the  tide, but on the primary's properties, and increases as $P_{tide}^{-2}$ or $P_{tide}^{-10/3}$ or $P_{tide} ^{-4}$ for a fast tide according to different laws \eqref{nu}. With \eqref{D_m} we readily obtain the tidal torque due to magnetic turbulence
\begin{equation}\label{Gamma_m}
\Gamma_m \approx -\beta_m\left(\frac{m}{M}\right)^2\left(\frac{R}{a}\right)^6R^5\frac{B^2}{\mu}(\omega_o-\omega_s)\frac{1}{\eta_t}.
\end{equation}
The structure factor $\beta_m=(16/9)\pi\int_\alpha^1 [x^{10}/(\int_0^x x'^2fdx')^2]dx$ $\approx 4.307$ for the present Sun. 

We still need to clarify some points. Firstly, it should be noted that tidal dissipation $D_m$ is the Ohmic dissipation on the tidal flow $u$, but not on the convective flow $v_c$. The tidal flow $u$ is much weaker than the convective flow $v_c$ in the binary systems we   discuss in the next section, such that tidal dissipation is much weaker than the total Ohmic dissipation in convection zone. Secondly, it should be noted that tidal torque $\Gamma_m\propto B^2$, such that a small change in magnetic field will lead to an appreciable change in tidal torque. Thirdly, if we apply the equipartition between magnetic and kinetic energies $B^2/\mu\approx\rho v_c^2$ and the estimation of convective velocity $v_c\approx(F/\rho)^{1/3}$, then we readily obtain $B^2/\mu\approx\rho^{1/3}F^{2/3}$. However, this equipartition based on the assumption of isotropic turbulence can be only applied to a slow rotator, but a young star rotates rapidly to break this assumption \citep{wei2022}. Therefore, for the application in the next section we do not use the equipartition, but instead the observed magnetic field.

\section{Application to the three systems}\label{sec:application}
In this section we apply our theory to  three binary systems: 51 Pegasi b, WASP-12b, and close binary stars in open clusters. The first two are planetary systems. We  make the estimations for the tidal evolution timescales of these systems.

\subsection{51 Pegasi b}
51 Pegasi b is a hot Jupiter, and is the first detected extroplanet orbiting a solar-type star \citep{mayor1995}. The short distance (0.05 AU) from the host star inhibits   planet formation because the high temperature of about 2000 K prevents gas accretion onto the  planet. Then this short distance is interpreted with the planet migration in the protoplanetary disk during the early evolution; in other words, the  angular momentum transfer in the disk  induces an inward migration. The question then arises of  what stops the inward migration of a planet and prevents it  from plunging into the host star. \citet{lin1996} proposed two mechanisms to interpret the stop of inward migration, namely tidal torque and magnetic torque. In the  mechanism of tidal torque, the standard equilibrium tide due to hydrodynamic turbulence is applied. Now we revisit the estimation of tidal torque due to either hydrodynamic or magnetic turbulence.

The rotation of a planet is already synchronized with the orbit due to its small moment of inertia, but the host star has not yet been synchronized with the orbit, and so we consider only the tidal torque raised by the planet on the host star. For a nearly circular orbit the migration timescale is
\begin{equation}\label{tau_a}
\tau_a=\frac{a}{|\dot a|}=\frac{L}{2|\Gamma|}.
\end{equation}
In Eq.  \eqref{tau_a} the orbital angular momentum $L=M_r(GM_ta)^{1/2}\approx m(GMa)^{1/2}$, where the total mass $M_t=M+m\approx M$ and the reduced mass $M_r=Mm/(M+m)\approx m$ for $m\ll M$. We take the parameters from the NASA website \footnote{https://exoplanetarchive.ipac.caltech.edu/}: $M=1.07 M_\odot$, $m=0.47 M_J$, $R=1.29 R_\odot$, and $a=0.05$ AU. At the young age we take a stellar rotation period of 1 day and a surface field $B=1000$ gauss (internal field 3500 gauss). We test the four models of turbulent viscosity $\nu_t$, namely $\nu_t=(1/3)v_c l$ in the absence of a fast-tide suppression effect on turbulent viscosity, as well as the three laws in Eq. \eqref{nu}. The quantities $v_c=35$ m/s and $l=10^7$ m are chosen for solar values, and turbulent magnetic diffusivity is assumed to be $\eta_t=\nu_t$. We estimate the two tidal torques $\Gamma_h$ and $\Gamma_m$ by \eqref{Gamma_h} and \eqref{Gamma_m}, and the two corresponding migration timescales $\tau_{ah}$ and $\tau_{am}$ by \eqref{tau_a}. We also calculate the tidal $Q_h$ due to hydrodynamic turbulence. Since tidal torque is inversely proportional to tidal $Q$, we can obtain the equivalent tidal $Q_m=(\Gamma_h/\Gamma_m)Q_h$ due to magnetic turbulence.

The calculation results are listed in Table \ref{table1}. First, as we have shown, with a stronger suppression effect on the turbulent viscosity or magnetic diffusivity, $\Gamma_h$ decreases but $\Gamma_m$ increases. With respect to the migration timescale, the tidal torque due to hydrodynamic turbulence is too weak, but the tidal torque due to magnetic turbulence with the  5/3 or square law can lead to migration timescales comparable to or less than several Myr (i.e., the viscous diffusion timescale of a protoplanetary disk) \citep{lin1996}, which indicates that the tidal torque due to magnetic turbulence is sufficiently strong to stop the inward migration of planet in the protoplanetary disk.
\begin{table*}
\centering
\begin{tabular}{c|c|c|c|c|c}
$\Gamma_h$ (N$\cdot$m) & $\Gamma_m$ (N$\cdot$m) & $Q_h$ & $Q_m$ & $\tau_{ah}$ (Myr) & $\tau_{am}$ (Myr) \\ \hline
0.16E+22 & 0.59E+26 & 0.25E+08 & 0.66E+03 & 0.93E+07 & 0.25E+03 \\ \hline
0.16E+21 & 0.58E+27 & 0.25E+09 & 0.67E+02 & 0.92E+08 & 0.25E+02 \\ \hline
0.15E+20 & 0.60E+28 & 0.26E+10 & 0.65E+01 & 0.95E+09 & 0.24E+01 \\ \hline
0.16E+19 & 0.57E+29 & 0.24E+11 & 0.68E+00 & 0.90E+10 & 0.26E+00
\end{tabular}
\caption{Calculation results for the  51 Pegasi b system. The stellar surface field is 1000 gauss. The first row is in the absence of fast-tide suppression effects on turbulent viscosity, the second row with linear law, the third row with 5/3 law, and the fourth row with square law.}\label{table1}
\end{table*}

\subsection{WASP-12b}
WASP-12b is the only hot Jupiter reported to have a decaying orbit on a timescale $P/\dot P\approx 3.2$ Myr \citep{maciejewski2011, patra2017, turner2021}. This, according to Kepler's third law, corresponds to $\tau_a=1.5 P/\dot P\approx 4.8$ Myr. Various models have been proposed to interpret the orbital decay, for example a companion star \citep{bechter2014}, apsidal precession \citep{maciejewski2016}, and tidal interaction in which dynamical tides due to nonlinear breaking of gravity waves can lead to such a short orbital decay timescale provided that the host star has evolved to the subgiant branch \citep{weinberg2017}. \citet{goodman2019} examined all the above models. The companion star and apsidal precession models cannot explain the short orbital decay timescale. For the dynamical tide model, the host star is not on the subgiant branch but on the main sequence. On the other hand, it is known that WASP-12b near the Roche limit of its host star is losing mass at a rate of about $10^{-7} M_J$ per year, and such a high mass loss rate probably has an appreciable impact on orbital decay because $M_p/\dot M_p\sim 2a/\dot a$. \citet{lin2010} proposed that the energy source of mass loss is internal tidal inflation due to orbital eccentricity $\approx 0.05$, but \citet{lai2010} pointed out that eccentricity is less than 0.02 and it is too small to lead to this mass loss rate. Since eccentricity cannot be precisely observed, we do not investigate mass loss induced by eccentricity, but consider a nearly circular orbit and try our model of magnetic turbulence to interpret the orbital decay.

We use the parameters from the  TESS observations \citep{turner2021}: $M=1.55 M_\odot$, $m=1.46 M_J$, $R=1.66 R_\odot$, and $a=0.024$ AU ($M$ and $a$ are combined to ensure the precisely observed $P=1.09$ days). The stellar rotation period is $\approx 30$ days. By $\log(L_*/L_\odot)=0.6$ we estimate $v_c\approx(F/\bar\rho)\approx 60$ m/s. Since the host star is a slow rotator at the age of $\approx 2$ Gyr, the surface field is taken to be 10 gauss. As we did for  51 Pegasi b, we estimate the tidal torques and migration timescales for WASP-12b with the four models of $\nu_t$.

The calculation results are listed in Table \ref{table2}. The tidal torque due to hydrodynamic turbulence is again too weak, but  due to magnetic turbulence it can lead to a sufficiently short $\tau_{am}$. The square law corresponds to $\tau_{am}$ shorter than 4.8 Myr. If we slightly reduce the surface field from 10 gauss to 8.4 gauss, then the tidal torque $\Gamma_m\propto B^2$ with the square law leads to $P/\dot P\approx$ 3.2 Myr.
\begin{table*}
\centering
\begin{tabular}{c|c|c|c|c|c}
$\Gamma_h$ (N$\cdot$m) & $\Gamma_m$ (N$\cdot$m) & $Q_h$ & $Q_m$ & $\tau_{ah}$ (Myr) & $\tau_{am}$ (Myr) \\ \hline
-0.78E+25 & -0.24E+26 &  0.14E+08 &  0.44E+07 &  0.48E+04 &  0.15E+04 \\ \hline
-0.12E+25 & -0.17E+27 &  0.94E+08 &  0.65E+06 &  0.33E+05 &  0.23E+03 \\ \hline
-0.14E+24 & -0.13E+28 &  0.76E+09 &  0.81E+05 &  0.26E+06 &  0.28E+02 \\ \hline
-0.17E+23 & -0.11E+29 &  0.63E+10 &  0.98E+04 &  0.22E+07 &  0.34E+01 \\ \hline
\end{tabular}
\caption{Similar to Table \ref{table1}, but for the  WASP-12b system. The stellar surface field is 10 gauss.}\label{table2}
\end{table*}

\subsection{Close binary stars in open clusters}
Close binary stars in open clusters are often used to test tidal theory because their ages can be robustly determined \citep{mayor1991, latham1992, mathieu1994, mathieu2005}. Observations show that binaries in young open clusters (less than a few hundred Myr) are circularized with orbital periods of  less than $P_{circ}\approx 8$ days, while old clusters ($>1$ Gyr) exhibit $P_{circ}\approx 15$ days \citep{wu2021}. Both equilibrium tides \citep{zahn1989, goodman1997, goodman1998} and dynamical tides \citep{goodman1998, savonije2002, weinberg2012, barker2020, wu2021} with resonance locking or nonlinear damping are applied to interpret the circularization timescale.  During the main sequence of binary stars, circularization timescales with orbital periods longer than 8 days cannot be interpreted via the standard theory of hydrodynamic equilibrium tides, and so the interpretation of tidal circularization during the pre-main sequence evolution was proposed \citep{goodman1997}. Now we describe  how we  interpret tidal circularization with our model of magnetic turbulence.

During the evolution of binary stars, the  orbital semimajor axis $a$ and eccentricity $e$ both decrease. The circularization timescale and the synchronization timescale differ by a factor, namely the ratio of orbital angular momentum to spin angular momentum $\tau_e/\tau_\omega\approx L_o/L_s$ \citep{zahn1989, ogilvie2014}, and this ratio scales as $L_o/L_s\propto (a/R)^2\gg 1$. Thus, binary stars are already  synchronized ($\omega_o\approx\omega_s$) long before circularization. Therefore, the two tidal bulges are along the line of two stellar centers such that tidal torque vanishes. However, orbital energy $E=-GMm/(2a)$ dissipates due to hydrodynamic or magnetic turbulence inside stars such that the semimajor axis $a$ decreases. In the absence of tidal torque, orbital angular momentum $L=M_r\left(GM_ta(1-e^2)\right)^{1/2}$ is conserved. Since $a$ decreases, $e$ should also decrease. This circularization mechanism with zero torque but energy dissipation was first proposed by Goldreich for satellite orbital circularization \citep{goldreich1963, goldreich1966}. With $\dot E=-D$ and $\dot L=0$, we readily obtain the circularization timescale
\begin{equation}\label{tau_e}
\tau_e=\frac{e}{|\dot e|}=\frac{|E|}{D}\left(\frac{2e^2}{1-e^2}\right).
\end{equation}
Circularization in the synchronized state is attributed to three eccentricity tides;   the dominant one has the tidal potential with its magnitude $\Psi=3.5e(Gmr^2/a^3)$ (the coefficient 3.5 is the magnitude ratio of this eccentricity tide to asynchronized tide), corresponding to the equilibrium tide $\xi=\Psi/g=3.5e(m/M(r))(r^4/a^3)$ and tidal flow $u=\dot\xi=\omega_o\xi$. Inserting $u$ into \eqref{D_h} and \eqref{D_m} we obtain disspations $D_h$ and $D_m$, and hence $\tau_{eh}$ and $\tau_{em}$ by \eqref{tau_e}. Since $D\propto u^2\propto e^2$, $\tau_e\propto 1/(1-e^2)$ becomes independent of $e$ at small $e$ (e.g., $e<0.3$), and we consider this situation to derive
\begin{equation}\label{tau_eam}
\tau_{eh}=\frac{1}{3.5^2\beta_h}\frac{a^8}{R^6}\frac{1}{\nu_t}, \hspace{3mm} \tau_{em}=\frac{1}{3.5^2\beta_m}M\frac{a^8}{R^{11}}\left(\frac{B^2}{\mu}\right)^{-1}\eta_t.
\end{equation}
Here we  reduce the circularization timescale by half provided that binary stars have comparable mass $M$.

The circularization timescale strongly depends on the  semimajor axis $a$, which decreases during the binary evolution and in turn influences $\nu_t$ and $\eta_t$ (see Eq. \eqref{nu}). On the other hand, the stellar magnetic field $B$ weakens during the stellar evolution. Since there are some uncertainties, we estimate the limits of the circularization timescale and calculate the three cases, namely tide due to hydrodynamic turbulence, and tide due to magnetic turbulence in solar-type binary stars at young ages with strong fields and at old ages with weak fields. We take solar values of  $M=M_\odot$, $R=R_\odot$, $v_c=35$ m/s, and $l=10^8$ m, and surface field $B=1000$ gauss at young ages and 10 gauss at old ages. We test the four models of fast-tide suppression effects as we did for the two planetary systems, and in Eq. \eqref{nu} the tidal period $P_{tide}=2\pi/\omega_o$. To compare with observations, we plot the circularization timescale versus the orbital period instead of the  semimajor axis used in \eqref{tau_eam}.

\begin{figure*}
\centering
\includegraphics[scale=0.35]{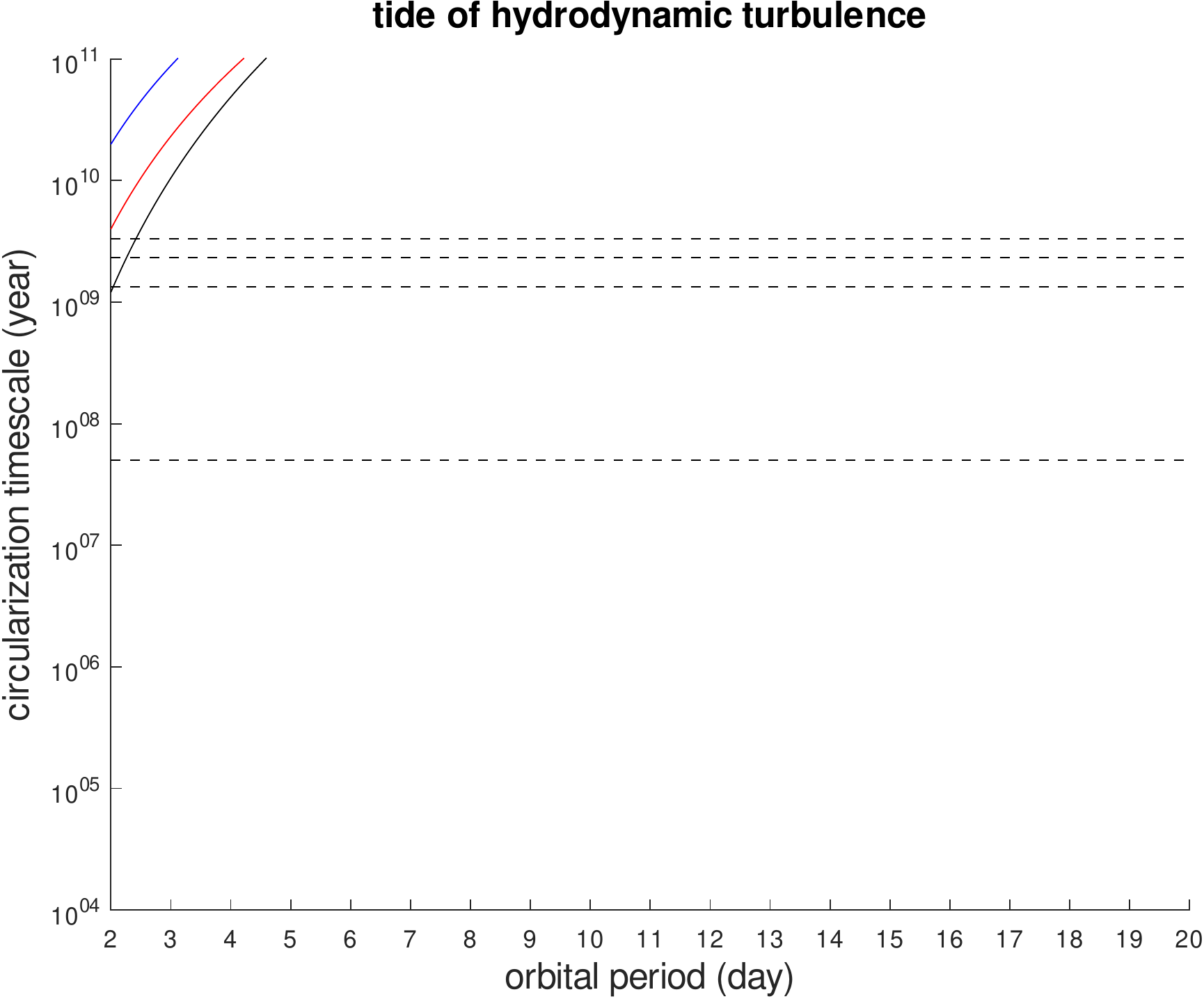}
\includegraphics[scale=0.35]{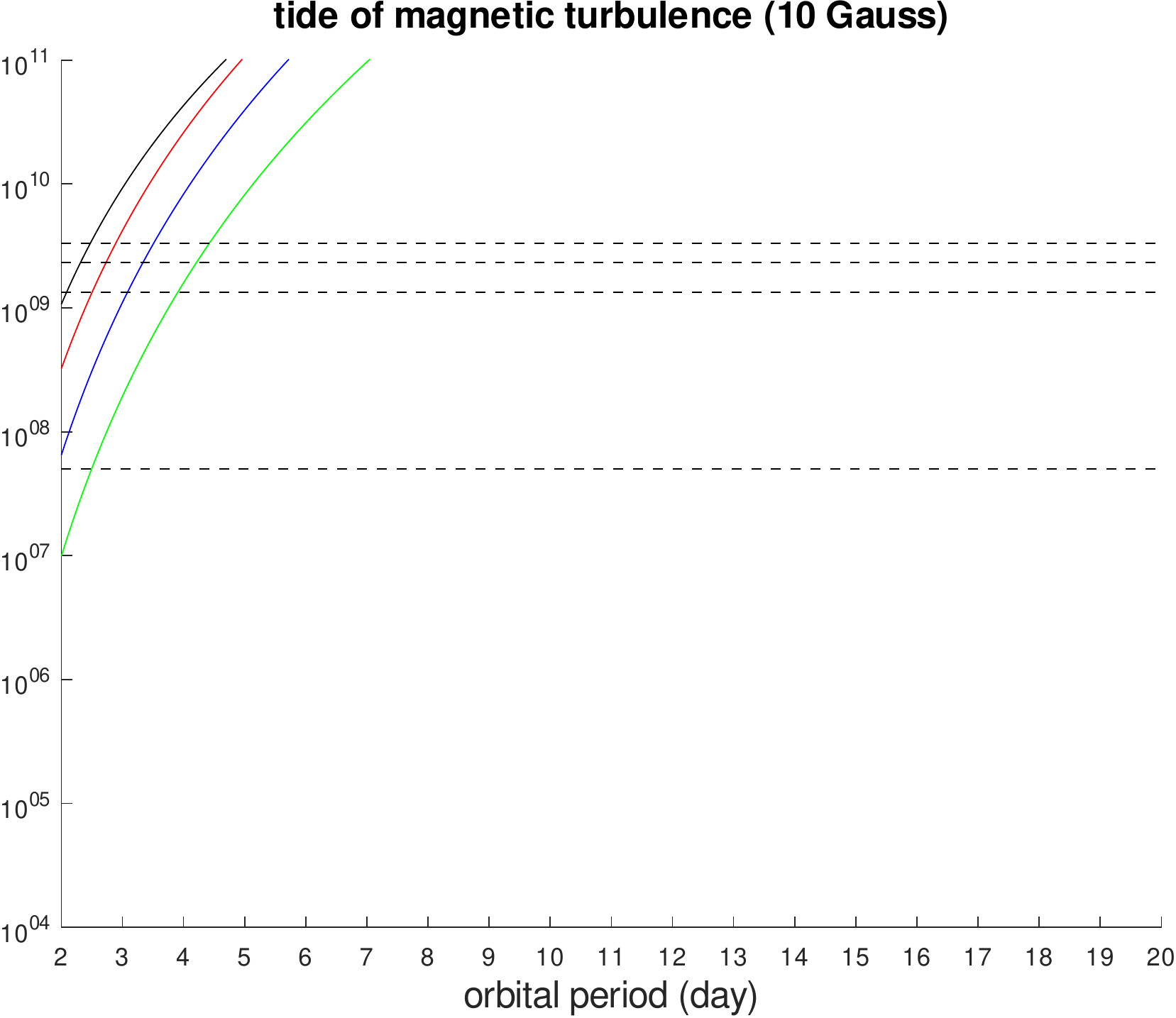}
\includegraphics[scale=0.35]{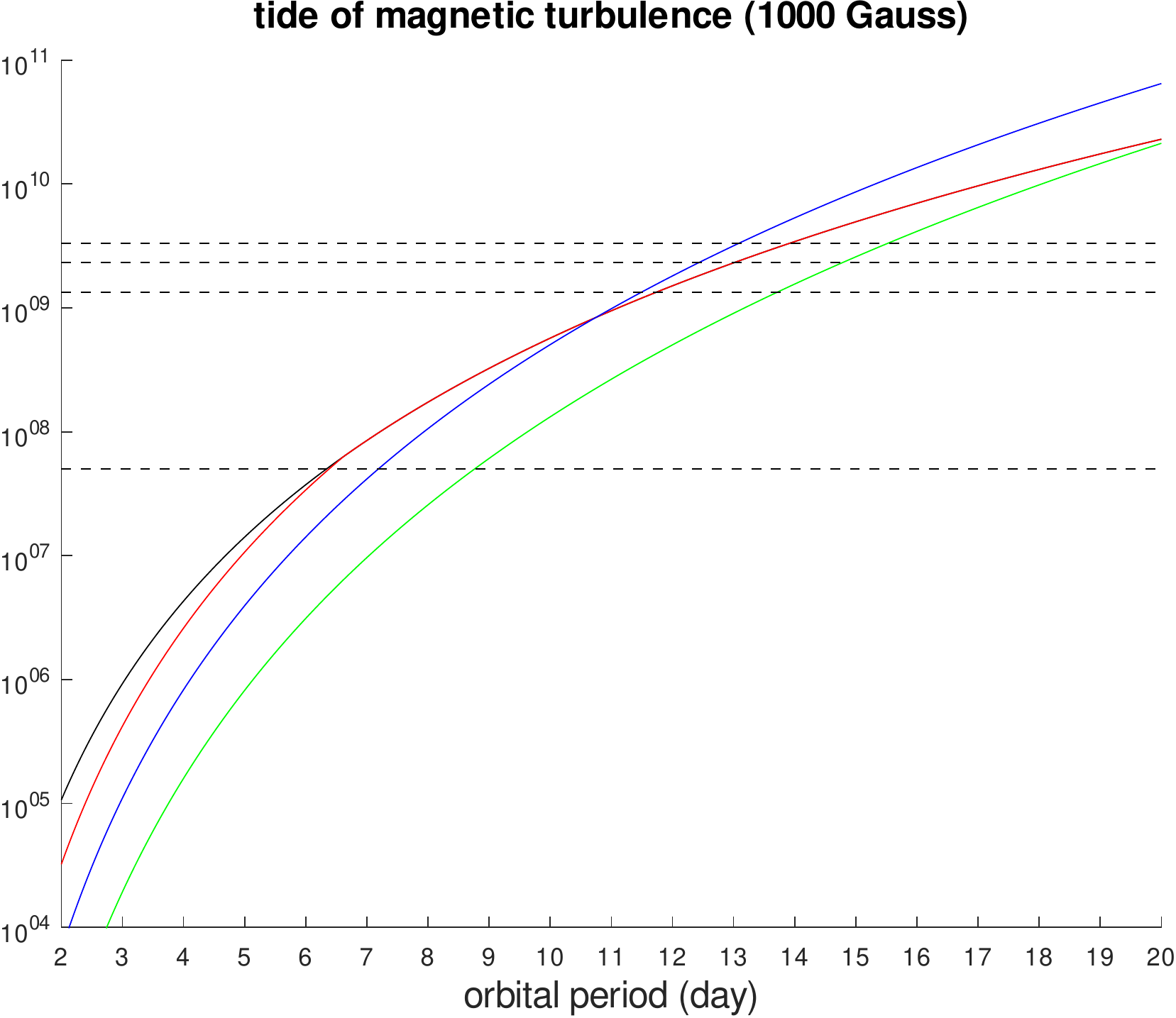}
\caption{Circularization timescale vs. orbital period in binary star systems. The black, red, blue, and green curves denote different laws of fast-tide suppression effects:  in absence of suppression, and with linear, 5/3, and square laws, respectively. The horizontal dashed lines from bottom to top denote one-third of the  age of the four open clusters from young to old: M35 of 150 Myr, M67 of 4 Gyr, NGC188 of 7 Gyr, and Halo of 10 Gyr.}\label{fig1}
\end{figure*}

The calculation results are shown in Figure \ref{fig1}. The horizontal dashed lines denote one-third of the age of the four open clusters: M35 of 150 Myr, M67 of 4 Gyr, NGC188 of 7 Gyr, and Halo of 10 Gyr. We take one-third of the age to ensure that during the entire life of the binary stars the orbital eccentricity $e=0.3$ decreases to $e=0.3\exp(-3)\approx 0.01$, which is sufficiently small to be considered   a circular orbit. For the young cluster M35, the tide due to hydrodynamic turbulence is too weak and $P_{circ}<2$ days. The tide due to magnetic turbulence with a surface field of 10 gauss corresponds to $P_{circ}<3$ days, but the  tide due to magnetic turbulence with a surface field of 1000 gauss can lead to $P_{circ}\approx 9$ days, which is  comparable to observations. For the old clusters, the first case corresponds to $P_{circ}<3$ days and the second case to $P_{circ}<5$ days, but the last case can lead to $P_{circ}>15$ days, again comparable to the observations. Therefore, we  infer that tides due to magnetic turbulence in the early evolution of binary stars with strong magnetic fields considerably contribute to orbital circularization. Recently, \citet{zanazzi2022} reported the observational results that there are two circularization periods, namely $P_{circ}\approx 3$ days for eccentric binaries and $P_{circ}>10$ days for circular binaries. This might be caused by the distribution of initial magnetic fields.

\section{Summary}\label{sec:summary}
We find that magnetic fields can significantly facilitate the tidal evolution of a binary system. This new physical mechanism arising from Ohmic dissipation rather than viscous dissipation leads to the order-of-magnitude change in the timescales of orbital evolution. Although in our estimations there are some uncertainties, for example  magnetic fields, structure factors, convective velocity, and mixing length, our theory provides a good approach to understanding the tidal evolution of binary systems.

\section*{Acknowledgments}
I thank Douglas N. C. Lin, Andrew Cumming, Xianfei Zhang and Tao Cai for their helpful discussions. This work is supported by National Natural Science Foundation of China (11872246, 12041301) and Beijing Natural Science Foundation (1202015).

\bibliographystyle{aa}
\bibliography{paper}

\appendix
\section{Estimation of electric current}
In this section we first  illustrate why the turbulent Ohmic dissipation is $\sim J^2/\sigma_t$. According to  the mean field theory for magnetohydrodynamic (MHD) turbulence \citep{moffatt1978}, the correlation of turbulent convective velocity $\bm v'$ and turbulent magnetic field $\bm B'$ (i.e., turbulent electromotive force) can be modeled with the large-scale field $\bm B$
\begin{equation}
\langle\bm v'\times\bm B'\rangle=\alpha\bm B-\beta\bm\nabla\times\bm B
,\end{equation}
where the coefficients $\alpha$ and $\beta$ are pseudo-tensors for anisotropic turbulence, but scalars for isotropic turbulence in our simple analysis.  The coefficient $\alpha$ is related to fluid helicity $\bm v'\times(\bm\nabla\times\bm v')$ and the terminology ``$\alpha$ effect'' in dynamo theory arises from this coefficient. Inserting the mean field theory into the magnetic induction equation we can derive magnetic diffusion to be $(\eta+\beta)\nabla^2\bm B$, where $\eta=1/(\sigma\mu)$ is the laminar magnetic diffusivity. We denote $\beta=\eta_t$ and in turbulence $\eta_t\gg\eta$ (i.e., magnetic diffusion $\approx\eta_t\nabla^2\bm B$). We then introduce turbulent electric conductivity $\sigma_t$ so that $\eta_t=1/(\sigma_t\mu)$. We now perform the dot product with $\bm B/(2\mu)$ on the magnetic induction equation
\begin{equation*}
\frac{\partial\bm B}{\partial t}=\bm\nabla\times(\bm v\times\bm B)+\eta_t\nabla^2\bm B
\end{equation*}
to derive magnetic energy equation. The Ohmic dissipation arises from the last term,
\begin{equation}\label{dissipation}
\frac{\bm B}{2\mu}\cdot(\eta_t\nabla^2\bm B)=-\frac{\eta_t}{2}\bm B\cdot(\bm\nabla\times\bm J)=\frac{\eta_t}{2}\left(\bm\nabla\cdot(\bm B\times\bm J)-\mu J^2\right)
,\end{equation}
where the vector identities $\nabla^2\bm B=-\bm\nabla\times(\bm\nabla\times\bm B)$ and $\bm\nabla\cdot(\bm B\times\bm J)=\bm J\cdot(\bm\nabla\times\bm B)-\bm B\cdot(\bm\nabla\times\bm J)$ are employed. In \eqref{dissipation} the first term $\bm\nabla\cdot(\bm B\times\bm J)$ almost vanishes for a volume integral over convection zone, and the second term is the turbulent Ohmic dissipation $\sim J^2/\sigma_t$.

Next we derive why the electric current $J$ associated with tide is $\sim\sigma_t u B$. The electric current $\bm J$ associated with tidal flow $\bm u$ is calculated by Ohm's law
\begin{equation}\label{J}
\bm J=\sigma_t(\bm E+\bm u\times \bm B+\bm u\times \bm b)
,\end{equation}
where $\bm E$ is the electric field associated with tidal flow, $\bm B$ is the mean field generated by dynamo in stellar convection zone, and $\bm b$ is the field induced by tidal flow $\bm u$. It should be noted that the mean field $\bm B$ is associated with a strong electric current in dynamo action, but what we are concerned with is not this current due to turbulent convection, but the current $\bm J$ associated with tidal flow $\bm u$. In this sense, $\bm B$ is assumed to be steady and potential for the sake of simplicity (it is possible to  involve the dynamo-generated current in the following analysis, but it is irrelevant to our calculation of tidal dissipation). To make the order-of-magnitude estimations, we introduce the length scale $l_b$ associated with the tidally induced field $\bm b$, which should be much less than the large-scale R, which is the  stellar radius. Ampere's law $\mu\bm J=\bm\nabla\times\bm b$ gives the estimation $J\simeq b/(\mu l_b)$. Faraday's law states $\bm\nabla\times\bm E=-\partial\bm b/\partial t$.  The electric field is caused by the tidal flow on the length scale $l_b$ and the timescale of $\bm b$ is tidal period $\tau$. Thus, Faraday's law yields the estimation $E\simeq bl_b/\tau$. We now estimate each term in Ohm's law \eqref{J}
\begin{equation}\label{order}
\begin{array}{cccc}
\textcircled{1}: \bm J, &\textcircled{2}: \sigma_t\bm E, &\textcircled{3}: \sigma_t\bm u\times \bm B, &\textcircled{4}: \sigma_t\bm u\times \bm b \\
b/(\mu l_b) &\sigma_t bl_b/\tau &\sigma_t uB &\sigma_t ub 
\end{array}
,\end{equation}
where the first row shows the four terms labeled with circled numbers and the second row shows the estimations. Then we build the balance among these terms. It is impossible that the four terms are at the same order, because \textcircled{3} and \textcircled{4} cannot be comparable since $b\ll B$ (i.e., the tide cannot be as strong as the convection). It is also impossible that the three terms are at the same order because the left term cannot be balanced. Thus, the only possibility is that two terms are in balance and the other two are in balance. Then there are   three combinations: $\textcircled{1}\simeq\textcircled{2}$ and $\textcircled{3}\simeq\textcircled{4}$; $\textcircled{1}\simeq\textcircled{3}$ and $\textcircled{2}\simeq\textcircled{4}$; $\textcircled{1}\simeq\textcircled{4}$ and $\textcircled{2}\simeq\textcircled{3}$. The first combination cannot hold since $b\ll B$ (i.e., $\textcircled{3}\simeq\textcircled{4}$ cannot hold). In the second combination, $\textcircled{2}\simeq\textcircled{4}$ leads to $l_b\simeq u\tau\simeq\xi$, and $\textcircled{1}\simeq\textcircled{3}$ leads to $b/B\simeq ul_b/\eta_t\simeq u\xi/\eta_t$. In the third combination, $\textcircled{1}\simeq\textcircled{4}$ leads to $\eta_t\simeq ul_b$, and $\textcircled{2}\simeq\textcircled{3}$ leads to $b/B\simeq u\tau/l_b\simeq\xi/l_b\simeq u\xi/\eta_t$. We now compare   these two combinations. Both   combinations yield $b/B\simeq u\xi/\eta_t$, which is reasonable because a stronger tide corresponds to a higher ratio $b/B$. However,  the difference between the two combinations is that $l_b\simeq\xi$ in the second combination, whereas $l_b\gg\xi$ (i.e., $b/B\simeq\xi/l_b$) in the third combination. By definition, $l_b$ is the length scale associated with the tidally induced field $\bm b$. Therefore, it is reasonable that $l_b$ should be comparable to tide $\xi$, but cannot be much larger than the  tide. Finally, we chose the second combination in which $J\simeq b/(\mu l_b)\simeq\sigma_t uB$.

\end{document}